\documentclass[twoside]{article}
\usepackage{fleqn,espcrc2}

\title{Linear and nonlinear realizations of superbranes}

\author{A. Kapustnikov\address{Department of Physics, Dnepropetrovsk University, \\
        49050, Dnepropetrovsk, Ukraine}\thanks{E-mail: kpstnkv@creator.dp.ua} and
        A. Shcherbakov\addressmark\thanks{E-mail: novel@ff.dsu.dp.ua}}

\begin{document}

\begin{abstract}
The coordinate transformations which establish the direct
relationship between the actions of linear and nonlinear
realizations of supermembranes are proposed. It is shown that the
Ro\v{c}ek-Tseytlin constraint known in the framework of the
linear realization of the theory is simply equivalent to  a limit
of  a "pure" nonlinear realization in which the field describing
the massive mode of the supermembrane puts to zero.
\end{abstract}
\maketitle

\section{Introduction}
One of the well-known examples of theory with partially broken
global supersymmetry ($PBGS$) is the
$N=1,~D=4$ supermembrane which can be derived either from the nonlinear
realization (NR) of global
$N=2$, $D=3$ supersymmetry partially broken down to $N=1$, $D=3$ or entirely
from a corresponding linear realization (LR) of this
supersymmetry \cite{ikr}. The first approach is more transparent.
It gives the manifestly covariant description of the action in
terms of vacuum Goldstone excitations associated with the
generators of the  spontaneously broken symmetries. The second one
is more simple but less transparent due to the presence of a bit
skillful nilpotency constraints suppressing the massive degree of
freedom of the supermembrane
\cite{rt}. In General these two approaches use different
Goldstone superfields and by now there is absent a manifestly
covariant procedure establishing the direct relation between them.

Here we would like to propose a new approach which establishes
the one to one correspondence between the quantities of linear
and nonlinear realizations of theory. The latter is based on a
special kind of superembedding of a complex superspace
${\bf C}^{3 \mid 2}$
into a superspace ${\bf C}^{4 \mid 2}$ in which the third spatial
coordinate is identified with a Goldstone superfield associated
with a central charge. The action of the supermembrane  appears
as a consequence of a nilpotency constraint
\cite{rt} imposed on the real part
of the Goldstone superfield. Note that because of this constraint
the massive mode of the supermembrane does not contribute into
the action. We show that from the geometrical point of view it
corresponds to a limit of a
{\bf pure nonlinear realization} in which the masses of all the
massive modes of theory tend to infinity.

\section{PBGS in complex superspace}
It was shown in \cite{ikr,pst} that the action of the $N=1,~ D=4$
supermembrane can be considered as the effective action of the
superfield theory with partial breaking global $N=2,~ D=3$
supersymmetry. This approach is based on the geometrical ideas of
Refs. \cite{va,hp,agit,git,bg} where the standard method of the NR of
supersymmetry in superspace were used. In this section we are going to
reproduce the basic entities of this
approach proceeding from the another prescription based on the ideas
of supergravity \cite{ik,ik1,ik2,os}.

\subsection{Linear realization}
Let us shortly consider the linear realization of the
$N=2,~D=3$ supersymmetry in the real superspace ${\bf R}^{3\mid
4}$
\begin{eqnarray} \label{RSSt}
 x^{\alpha \beta \prime} &=& {x^{\alpha \beta}}-\frac{i}{4}(\varepsilon^{\alpha}\tau^{\beta}+
 \xi^{\alpha}\omega^{\beta}+
      \alpha \leftrightarrow \beta),\\
 \tau^{\alpha \prime} &=& \tau^{\alpha} + \varepsilon^{\alpha},\quad
 \omega^{\alpha \prime} = \omega^{\alpha} + \xi^{\alpha},\nonumber
\end{eqnarray}
where $\alpha$, $\beta$ are the spinor indices of the Lorentz
group $SL(2,R)$. This superspace can be regarded as the real
subspace of the complex superspace ${\bf C}^{3\mid2}$
\begin{eqnarray} \label{CSS}
 { x_L^{\alpha \beta}}&=&{x^{\alpha \beta}}+\frac{1}{4}(\tau^{\alpha} \omega^{\beta} + \alpha \leftrightarrow \beta),\\
 {\theta^{\alpha}}&=&\frac{1}{\sqrt{2}}(\tau^{\alpha}+i\omega^{\alpha}),\nonumber
\end{eqnarray}
where the original supergroup is realized as follows
\[
 x_L^{\alpha \beta\prime}={ x_L^{\alpha \beta}}-\frac{i}{2}({\bar{\varepsilon}}^{\alpha}_{c} \theta^{\beta} +
 {\bar{\varepsilon}}^{\beta}_{c} \theta^{\alpha}) +
 \frac{i}{4}(\varepsilon^{\alpha}_{c} {\bar{\varepsilon}}^{\beta}_{c} +
 \varepsilon^{\beta}_{c} {\bar{\varepsilon}}^{\alpha}_{c}),\]
\begin{equation} \label{CSSt}
 {\theta^{\alpha}}{}^{\prime}={\theta^{\alpha}}+\varepsilon^{\alpha}_{c},
\end{equation}
\[ \varepsilon^{\alpha}_{c}=\frac{1}{\sqrt{2}}(\varepsilon^{\alpha} + i\xi^{\alpha}),\quad
 {\bar{\varepsilon}}^{\alpha}_c = \frac{1}{\sqrt{2}}(\varepsilon^{\alpha} -
 i\xi^{\alpha}).\]
To provide the partial breaking of this supersymmetry let us
consider the model of superembedding of the original superspace
${\bf C}^{3\mid2}$ into an extended complex superspace ${\bf
C}^{4\mid2}$ with an additional bosonic coordinate
$q$
\begin{eqnarray} \label{cct}
 q^{\prime} &=&
 q - \frac{1}{2}({\bar{\varepsilon}}^{\alpha}_{c} \theta_{\alpha} -
 {\varepsilon}_c^{\alpha}{\bar{\theta}}_{\alpha}),\\
 q_L^{\prime} &=& q_L - {\bar{\varepsilon}}^{\alpha}_{c} \theta_{\alpha} -
 \frac{1}{2}{\bar{\varepsilon}}^{\alpha}_c{\varepsilon}_{c\alpha},~~~
 q_L = q - \frac{1}{2}\theta^{\alpha}{\bar{\theta}}_{\alpha}. \nonumber
\end{eqnarray}
Eq.(\ref{cct}) together with (\ref{RSSt}) describes the
transformations of $N=2$, $D=3$ supersymmetry with a central
charge Z and the following generators of supertranslations
\[ \check{S}_{\alpha} = \frac{\partial}{\partial \tau^{\alpha}} -
 \frac{i}{2}\tau^{\beta}
 \frac{\partial}{\partial x^{\alpha\beta}} -
 \frac{1}{2}\omega_{\alpha}Z \equiv S_{\alpha} - \frac{1}{2}\omega_{\alpha}Z,\]
\[ \check{Q}_{\alpha} = \frac{\partial}{\partial \omega^{\alpha}} -
 \frac{i}{2}\omega^{\beta}
 \frac{\partial}{\partial x^{\alpha\beta}} +
 \frac{1}{2}\tau_{\alpha}Z \equiv Q_{\alpha} +
 \frac{1}{2}\tau_{\alpha}Z,\]
\begin{equation} \label{extG}
 Z \equiv i\frac{\partial}{\partial q}.
\end{equation}
To avoid the presence of the redundant coordinate $q$ in the
theory we impose the following superembedding condition
\begin{equation} \label{sem}
q_L = i\xi^{-1}\phi(x_L,\theta).
\end{equation}
It deserves mentioning that in accordance with the definition
(\ref{sem}) the superfield $\phi(x_L,\theta)$ transformes
inhomogeneously with respect to the transformations (\ref{cct})
 \begin{equation} \label{phitr}
 \phi(x_L^{\prime},\theta^{\prime}) = \phi(x_L,\theta) +
 i\xi{\bar{\varepsilon}}^{\alpha}_{c} \theta_{\alpha} +
 \frac{i\xi}{2}{\bar{\varepsilon}}^{\alpha}_c{\varepsilon}_{c\alpha}.
 \end{equation}
Therefore (\ref{phitr}) can be assumed to be a  Goldstone
superfield of
$N=2$, $D=3$ supersymmetry. To justify this assumption
it is instructive to notice that the real and imaginary parts of
the superfield
$\phi(x_L,\theta) \equiv {\mbox{Re}} \phi_L(x,\tau, \omega) + i{\mbox{Im}} \phi_L(x,\tau, \omega)$
satisfy the chirality conditions
\begin{eqnarray} \label{shcd}
D^{(\tau)}_{\alpha}{\mbox{Re}} \phi_L &=& D^{(\omega)}_{\alpha}{\mbox{Im}} \phi_L,\nonumber\\
D^{(\omega)}_{\alpha}{\mbox{Re}} \phi_L &=& - D^{(\tau)}_{\alpha}{\mbox{Im}}
\phi_L,\\
D^{(\tau)}_{\alpha} = \frac{\partial}{\partial \tau^{\alpha}} &+&\frac{i}{2}\tau^{\beta}\frac{\partial}{\partial x^{\alpha\beta}},\nonumber\\
D^{(\omega)}_{\alpha} = \frac{\partial}{\partial \omega^{\alpha}} &+&\frac{i}{2}\omega^{\beta}\frac{\partial}{\partial x^{\alpha\beta}}.\nonumber
\end{eqnarray}
This means that if we introduce the shifted superfield
\begin{equation} \label{shSF}
\phi_{\xi}(x,\tau, \omega) = \phi_L(x,\tau, \omega) - \frac{i\xi}{4}(\tau^2
+ \omega^2),
\end{equation}
which transforms as the real variable $q$
\begin{equation} \label{shSFtr}
\phi_{\xi}^{\prime}(x^{\prime},\tau^{\prime}, \omega^{\prime}) =
\phi_{\xi}(x,\tau, \omega) - \frac{\xi}{2}(\varepsilon^{\alpha}\omega_{\alpha} -
 \xi^{\alpha}\tau_{\alpha})
\end{equation}
and on which the central charge generator $Z$ is realized as
\begin{equation} \label{Zreal}
Z\phi_{\xi} = - \xi,\quad (Z\bar{\phi}_{\xi} = \xi),
\end{equation}
we reveal that its real and imaginary parts would satisfy the
same chirality conditions as the original superfield does but
with the modified spinor derivatives of the extended supersymmetry
\begin{equation} \label{lcd}
\check{D}^{(\tau)}_{\alpha} = D^{(\tau)}_{\alpha} +
\frac{1}{2}\omega_{\alpha}Z, \quad
\check{D}^{(\omega)}_{\alpha} = D^{(\omega)}_{\alpha} - \frac{1}{2}\tau_{\alpha}Z.
\end{equation}
Thus the superfield $\phi_{\xi}$ is also chiral (but in contrast
to $\phi_L$) with respect to the extended supersymmetry. Another
very important feature of the superfield $\phi_{\xi}$ is that its
real part possesses the homogeneous transformation law under the
both of supersymmetries. The careful analysis shows that
$\phi_{\xi}$ indeed insures the partial breaking of
$N=2$, $D=3$ supersymmetry and one of its spinor $N=1$ components describes
the Goldstone excitations of the $4D$ supermembrane associated with
$Q$-supertranslations. Let us write down the
Grassmann decomposition of the original chiral superfield (\ref{sem})
\begin{eqnarray} \label{chSFLRcomp}
{\mbox{Re}} \phi_L &=& A + i\omega^{\alpha}\Lambda_{\alpha} -
\frac{i}{4}\omega^2F, \\
{\mbox{Im}} \phi_L &=& B + i\omega^{\alpha}\Sigma_{\alpha} -
\frac{i}{4}\omega^2G. \nonumber
\end{eqnarray}
Substituting (\ref{chSFLRcomp}) into the chirality conditions
(\ref{shcd}) and solving them one gets
\begin{eqnarray} \label{ABdec}
 {\mbox{Re}} \phi_L = A -
 \omega^{\alpha}D^{(\tau)}_{\alpha}B +
 \frac{1}{4} \omega^2 D^{(\tau) \alpha}D^{(\tau)}_{\alpha}A,&&\\
 {\mbox{Im}} \phi_L = B + \omega^{\alpha} D^{(\tau)}_{\alpha} A +
 \frac{1}{4} \omega^2 D^{(\tau) \alpha}D^{(\tau)}_{\alpha}B. &&\nonumber
\end{eqnarray}
This leads to the following solution for the $N=1$ components of
the shifted superfield
\begin{eqnarray} \label{ABxidec}
 {\mbox{Re}} \phi_{\xi} &=& A - \frac{i\xi}{4}\tau^2 -
 \omega^{\alpha}D^{(\tau)}_{\alpha}B \nonumber\\
 &&+ \frac{1}{4} \omega^2 (-i\xi + D^{(\tau) \alpha}D^{(\tau)}_{\alpha}A),\\
 {\mbox{Im}} \phi_{\xi} &=& B + \omega^{\alpha} D^{(\tau)}_{\alpha} A +
 \frac{1}{4} \omega^2 D^{(\tau) \alpha}D^{(\tau)}_{\alpha}B. \nonumber
\end{eqnarray}
It is not hard to verify that the superfield
$\Lambda_{\alpha} = iD^{(\tau)}_{\alpha}B$
transforms inhomogeneously under the second supersymmetry
$(\delta = -\xi^{\alpha}Q_{\alpha})$
\[
\delta \Lambda_{\alpha} = \xi \xi_{\alpha} +
\frac{i}{2}\xi_{\alpha}D^{(\tau) \beta}D^{(\tau)}_{\beta}A_{\xi} -
\frac{1}{2}\xi^{\beta}\partial_{\alpha\beta}A_{\xi},\]
\begin{equation} \label{GFLR}
A_{\xi} = A - \frac{i\xi}{4}\tau^2,
\end{equation}
because of the transformation law of the Goldstone superfield $B$
associated with the central charge
\begin{equation} \label{GBLR}
\delta B = - i\xi - \xi^{\alpha}D^{(\tau)}_{\alpha}A_{\xi}.
\end{equation}
Thus we conclude that the superfields $\Lambda_{\alpha}$ and $B$
actually represent the Goldstone excitations of a theory with
partially broken $N=2$, $D=3$ supersymmetry.

Now we can directly follow the prescriptions of Ref.\cite{rt} and
impose the covariant constraint
\begin{equation} \label{RTcon}
({\mbox{Re}}\phi_{\xi})^2 = 0,
\end{equation}
which allows one to eliminate the massive mode of supermembrane $A$
in terms of the Goldstone superfield
$\Lambda_{\alpha}$
\begin{equation} \label{A}
A_{\xi} \equiv A - \frac{i}{4}\xi\tau^2 =
 \frac{-i\Lambda^2}{\xi + \sqrt{\xi^2 + (D^{(\tau)})^2\Lambda^2}}.
\end{equation}
This equation coincides with that obtained in \cite{ikr}. The
difference is in the method of its derivation. In our approach it
was obtained in the frame of the linear realization of
supersymmetry after imposing the nilpotency constraint
(\ref{RTcon}), while in
\cite{ikr} the starting point was the nonlinear realization and
the constraint (\ref{RTcon}) was not exploited at all. The reason
of this
discrepancy is that in the NR method 
of PBGS the massive mode of the supermembrane and the
corresponding Goldstone superfield are transformed independently
with respect to both supersymmetries. Moreover there always
exists the possibility of avoiding  the massive mode of the
supermembrane by the putting to zero the corresponding $N=1$
superfield without violating the covariant properties of the $N=2$
chiral superfield composed of this mode and the related Goldstone
superfield. Further we shall prove that both of these approaches
are canonically equivalent to each other. More precisely, it will
be shown that there exists the change of the coordinates involved
which
transforms the superfield action in the LR method 
of the $4D$
supermembrane
\begin{eqnarray} \label{act}
S &=& i\int d^3xd^2\tau A_{\xi}(x,\tau) \nonumber\\
&\equiv& -\frac{i}{4}\int d^3xd^2\tau d^2\omega \omega^2{\mbox{Re}} \phi_{\xi},
\end{eqnarray}
into the related action of the NR method 
and vice versa if the
constraint (\ref{RTcon}) is replaced by the equivalent manifestly covariant
constraint of the ``pure'' NR formulation

\subsection{Unusual form of the nonlinear realization}
We have already seen that the chiral superspace ${\bf C}^{3\mid2}$
is the most suitable basis for describing models with partial
breaking of $N=2,~~D=3$ supersymmetry. It seems therefore
naturally to introduce the Goldstone superfields of the
underlying nonlinear realization
$\psi_{\alpha}(\tilde x,{\tilde \tau})$ and $Q(\tilde x,{\tilde \tau})$
directly in this basis. This is a very essential point of our
approach because following the chirality principle we are forced
to begin with transformations of a little bit unusual form
\[
{{\tilde x_L}^{\alpha \beta}}{}^{\prime}={{\tilde x_L}^{\alpha \beta}} - \frac{i}{4}(\varepsilon^{\alpha} {\tilde \theta}^{\beta} +
      \xi^{\alpha}\psi^{\beta}({\tilde x_L},{\tilde \theta}) + \alpha \leftrightarrow \beta),\]
\begin{equation}\label{NR}
\psi^{\alpha \prime}({\tilde x_L}{}^{\prime},{\tilde \theta}{}^{\prime}) = \psi^{\alpha}({\tilde x_L},{\tilde \theta})+
      \xi^{\alpha}.
\end{equation}
\[
{{\tilde \theta}^{\alpha}}{}^{\prime} = {{\tilde \theta}^{\alpha}} + \varepsilon^{\alpha},
\]
Note that instead original form \cite{ikr} of the nonlinear
realization, which is compatible with the transformations of the
coordinates in real superspace (\ref{RSSt}), this one is based on
the nonlinear transformations of the Goldstone superfields
$\psi_{\alpha}(\tilde x,{\tilde \tau})$ and $Q(\tilde x,{\tilde \tau})$ in
the complex superspace
${\bf \tilde{C}}^{3 \mid 2} = \{\tilde x,{\tilde \tau}\}$.
Below we will see that although this form of the nonlinear
realization looks quite unusual at first glance it is equivalent
to the ordinary one and transforms into it upon the canonical
redefinition of coordinates. But let us first consider some
unconventional properties of this nonlinear realization. It is
quite easy to check that the superfunctions
\[
 Y^{\alpha\beta}_L({\tilde x_L},{\tilde \theta}) = {{\tilde x_L}^{\alpha \beta}} +
 \frac{1}{4} \left( {{\tilde \theta}^{\alpha}} \psi^{\beta}({\tilde x_L},{\tilde \theta}) +
  \alpha \leftrightarrow \beta \right),\]
\begin{equation} \label{csSFS}
 Y^{\alpha}_L({\tilde x_L},{\tilde \theta}) = \frac{1}{\sqrt{2}} ({{\tilde \theta}^{\alpha}} +
 i\psi^{\alpha}({\tilde x_L},{\tilde \theta})).
\end{equation}
are transformed under (\ref{NR}) as the coordinates of the
superspace
${\bf C}^{3 \mid 2}$ in eq. (\ref{CSSt}). Therefore we can
suggest the transformations
\begin{equation} \label{map}
 { x_L^{\alpha \beta}} = Y^{\alpha\beta}_L({\tilde x_L},{\tilde \theta}), \quad
 {\theta^{\alpha}} = Y^{\alpha}_L({\tilde x_L},{\tilde \theta}),
\end{equation}
which establish the map
${\bf C}^{3 \mid 2} \Rightarrow {\bf \tilde{C}}^{3 \mid 2}$.
Our next step resembles the geometrical approach to supergravity
by Ogievetsky and Sokatchev \cite{os}. We consider the embedding
of a real superspace ${\bf R}^{3 \mid 4}$ into the complex one
${\bf C}^{3 \mid 2}$. This embedding as we know is described by the relation
(\ref{CSS}). If we take into account the coordinate
transformations (\ref{csSFS}) and (\ref{map}) we find a new
nontrivial axial-vector object
\begin{equation} \label{PrDef}
 {{\tilde x_L}^{\alpha \beta}} = {\tilde x^{\alpha \beta}} + {{\tilde H}^{\alpha \beta}}(\tilde x,{\tilde \tau},{\tilde \omega}),\quad {{\tilde \theta}^{\alpha}} = {\tilde \tau}^{\alpha} + i{\tilde \omega}^{\alpha}
\end{equation}
which provides a suitable map 
${\bf R}^{3 \mid 4} \Rightarrow {\bf \tilde{R}}^{3 \mid 4}$. Note that in
accordance with (\ref{NR}) the real spinor variable ${\tilde
\omega}^{\alpha}$ entering eqs. (\ref{PrDef}) as the
imaginary part of the complex spinor ${{\tilde \theta}^{\alpha}}$
does not transform at all with respect to both supersymmetries.
This fact reflects the idea of spontaneous breaking of global
supersymmetry in its pure geometrical form. The explicit form of
the superfield
${{\tilde H}^{\alpha \beta}}(\tilde x,{\tilde \tau},{\tilde \omega})$
as well as the expressions of the ``old'' real variables
$\{x, \tau, \omega \}$ through the ``new'' ones $\{\tilde x,{\tilde \tau},{\tilde \omega} \}$
can be read out straightforwardly from the Eqs. (\ref{CSS}),
(\ref{csSFS}) and (\ref{map}). Here we shall only present the
final results of these calculations
\begin{equation} \label{Pr}
 {{\tilde H}^{\alpha \beta}} (\tilde x,{\tilde \tau},{\tilde \omega}) = (i{\tilde \omega}^{\rho} T_{\rho}{}^{\mu\nu}  -
  \frac{1}{2}{\tilde \omega}^2 {\cal D}^{\mu} \psi^{\nu}) T^{-1}_{\mu \nu}{}^{\alpha \beta},
\end{equation}
\begin{eqnarray} \label{CVx}
 x^{\alpha \beta} &=&
 {\tilde x^{\alpha \beta}} + \frac{i}{4} \left( {\tilde \omega}^{\alpha}\psi^{\beta}
+ \tilde{\tau}^{\alpha}({\tilde \omega}^{\rho}{\cal
D}_{\rho}\psi^{\beta}\right.\\
&+& \frac{i{\tilde \omega}^2}{2}{\cal D}^{\rho}\psi^{\sigma}{\cal D}_{\rho \sigma}\psi^{\beta}) +
 \alpha \leftrightarrow \left. \beta \right),
\end{eqnarray}
\begin{equation} \label{CVtau}
 \tau^{\alpha} =
  {{\tilde \tau}^{\alpha}} - {\tilde \omega}^{\rho} {\cal D}_{\rho} \psi^{\alpha} - \frac{i{\tilde \omega}^2}{2}{\cal D}^{\rho}\psi^{\sigma}{\cal D}_{\rho
\sigma}\psi^{\alpha},
\end{equation}
\begin{equation} \label{CVom}
 \omega^{\alpha} =
  {\tilde \omega}^{\alpha} + \tilde{\psi}^{\alpha}(\tilde x, {\tilde \tau}, {\tilde \omega}),
\end{equation}
\begin{equation} \label{tpsi}
\tilde{\psi}^{\alpha}(\tilde x, {\tilde \tau}, {\tilde \omega}) = \psi^{\alpha}(\tilde x,{\tilde \tau})+
 \frac{{\tilde \omega}^2}{4}\tilde{\cal D}^2 \psi^{\alpha}(\tilde x,{\tilde \tau}),
\end{equation}
where
\begin{eqnarray} \label{NRent}
 {\cal D}_{\rho} &=& \tilde{\partial}_{\rho} + T_{\rho}{}^{\mu\nu}{\cal
 D}_{\mu \nu}\nonumber \\
 &=& D^{(\tilde{\tau})}_{\rho} + \frac{i}{4}(\psi^{\mu}D^{(\tilde{\tau})}_{\rho}\psi^{\nu} +
 \mu \leftrightarrow \nu){\cal D}_{\mu \nu},\nonumber\\
 {\cal D}_{\mu \nu} &=& T^{-1}_{\mu \nu}{}^{\alpha\beta}{\tilde \partial}_{\alpha
 \beta},\\
 \tilde{\cal D}^2 & \equiv & {\cal D}^{\beta}{\cal D}_{\beta} -
 ({\cal D}^{\beta}(TT^{-1})_{\beta}^{\mu\nu}) {\tilde \partial}_{\mu \nu} \nonumber
\end{eqnarray}
\[
 \tilde{\partial}{}_{\alpha \beta} = \frac{\partial}{\partial{\tilde{x}^{\alpha
 \beta}}},\quad
 D^{(\tilde{\tau})}_{\alpha} =
 \frac{\partial}{\partial \tilde{\tau}^{\alpha}} +
 \frac{i}{2}\tilde{\tau}^{\beta}\frac{\partial}{\partial{\tilde{x}^{\alpha \beta}}},
\]
\begin{equation} \label{NRenti}
 T_{\mu}{}^{\alpha\beta} = \frac{i}{4}({{\tilde \tau}^{\alpha}} \delta^{\beta}_{\mu} +
 \psi^{\alpha} {\tilde \partial}_{\mu} \psi^{\beta}
  + \alpha \leftrightarrow \beta),
\end{equation}
\[ T^{-1}_{\mu\nu}{}^{\alpha \beta} = \delta^{\alpha \beta}_{\mu \nu} -
 \frac{i}{4}(\psi^{\alpha} {\cal D}_{\mu \nu}
  \psi^{\beta} + \alpha \leftrightarrow \beta),\]
are the vielbein and covariant derivatives of the $N=2$, $D=3$
NR 
obtained in \cite{ikr}. Thus we see that the intrinsic geometry
of the $N=2$, $D=3$ PBGS theory is closely related to the axial
vector prepotential
$\tilde{H}^{\alpha \beta}(\tilde x, {\tilde \tau}, {\tilde \omega})$ which involves all
the basic ingredients  of the corresponding NR method. It deserves
mentioning that this example is a generalization of the
geometrical approach to supergravity proposed in
\cite{ik1} for the models with completely broken {\it local~}
supersymmetries. It was shown that in these cases the transition
to the complex superspace also gives the nontrivial axial-vector
prepotential which contains in its {\it flat limit~} the
covariant objects of the Volkov-Akulov nonlinear realization
\cite{va}. It is naturally to assume, that
eq.(\ref{Pr}) describes the {\it flat limit~} of the prepotential
related to
$N=2$, $D=3$ supergravity with partially broken {\it local~}
supersymmetry. We are not ready to consider this problem here to
full extent as it raise questions which go beyond the framework
of this report, and will proceed with the discussion of the
superbrane theory with rigid worldvolume supersymmetry.

\subsection{Connection with the usual form}
Let us note that the restriction of the map
${\bf C}^{3 \mid 2} \Rightarrow {\bf \tilde{C}}^{3 \mid 2}$ onto the real
superspace (\ref{map}), (\ref{PrDef}) leads to the following unconventional form of
the NR 
\[
 {\tilde x^{\alpha \beta}}{}^{\prime} = {\tilde x^{\alpha \beta}} - \frac{i}{4} \big[ \varepsilon^{\alpha} {\tilde \tau}^{\beta}
      + \xi^{\alpha} \tilde{\psi}^{\beta}(\tilde x, {\tilde \tau}, {\tilde \omega}) +
      \alpha \leftrightarrow \beta \big],\]
\begin{equation} \label{uncfNR}
 {\tilde \tau}^{\alpha \prime} = {\tilde
 \tau}^{\alpha}+\varepsilon^{\alpha},\quad {\tilde \omega}^{\alpha \prime} = {\tilde \omega}^{\alpha},
\end{equation}
$ \tilde{\psi}^{\alpha \prime}(\tilde x^{\prime},{\tilde \tau}^{\prime}, \omega^{\prime}) =
 \tilde{\psi}^{\alpha}(\tilde x, {\tilde \tau}, {\tilde \omega}) + \xi^{\alpha}. $

This form, however, is replaced by the usual one
\[ \hat{x}^{\alpha\beta\prime} = \hat{x}^{\alpha\beta} -
 \frac{i}{4} \big[ \varepsilon^{\alpha}\hat{\tau}^{\beta}
      + \xi^{\alpha} \psi^{\beta}(\hat{x}, \hat{\tau}) +
      \alpha \leftrightarrow \beta \big],\]
\begin{equation} \label{usfNR}
 \hat{\tau}^{\alpha \prime} =
 \hat{\tau}^{\alpha}+\varepsilon^{\alpha},\quad  \hat{\omega}^{\alpha \prime} = \hat{\omega}^{\alpha},
\end{equation}
$ \psi^{\alpha \prime}(\hat{x}^{\prime},\hat{\tau}^{\prime}) =
 \psi^{\alpha}(\hat{x}, \hat{\tau}) + \xi^{\alpha},$\\
upon a redefinition of the coordinates
\begin{equation} \label{hattildred}
\hat{z} = \{\hat{x}, \hat{\tau}, \hat{\omega} \} \Rightarrow
\tilde{z} = \{\tilde{x}, \tilde{\tau}, \tilde{\omega} \}
\end{equation}
To get the explicit form of the transformations
(\ref{hattildred}) one should change the variables as follows
\begin{eqnarray} \label{ordhatred}
x^{\alpha\beta}  &=& \hat{x}^{\alpha\beta} +
\frac{i}{4}[\hat{\omega}^{\alpha}\psi^{\beta}(\hat{x}, \hat{\tau}) +
      \alpha \leftrightarrow \beta],\\
\tau^{\alpha} &=& \hat{\tau}^{\alpha},~~~\omega^{\alpha} =
\hat{\omega}^{\alpha} + \psi^{\alpha}(\hat{x}, \hat{\tau}),\nonumber
\end{eqnarray}
which provides the usual form of the nonlinear realization
(\ref{usfNR})
\footnote{One can check that when applied to the r.h.s. of the
Eqs. (\ref{ordhatred}) the transformations (\ref{usfNR}) imply the
ordinary transformation laws of the variables
$\{x, \tau, \omega\}$.},
and then inverting the transformations (\ref{ordhatred})
\begin{eqnarray}\label{hatlin}
\hat{x}^{\alpha \beta} &=& x^{\alpha \beta} - \frac{i}{4} ( \omega^{\alpha} \psi^{\beta}\nonumber\\
 &+&\frac{i}{4} \omega^2 \varepsilon^{\alpha \mu} \psi^{\nu} \partial_{\mu \nu} \psi^{\beta} +
 \alpha
\leftrightarrow \beta ),
\end{eqnarray}
\[\hat{\omega}^{\alpha} = \omega^{\alpha} - \psi^{\alpha} + \frac{i}{2} \omega^{\mu} \psi^{\nu}
\partial_{\mu \nu} \psi^{\alpha}\]
\[ -\frac{1}{8} \omega^2 \varepsilon^{\mu \rho} \psi^{\sigma} (
\partial_{\rho
\sigma} \psi^{\nu} \partial_{\mu \nu}\psi^{\alpha} +
\frac{1}{2} \psi^{\nu} \partial_{\mu \nu} \partial_{\rho \sigma}
\psi^{\alpha} ),\]
\[\hat{\tau}^{\alpha} = \tau^{\alpha}\]
by substituting into (\ref{ordhatred}) the superfunctions
(\ref{CVx}), (\ref{CVtau}), (\ref{CVom}) instead of the variables
$z = \{x,
\tau, \omega \}$.

\section{The limit of pure nonlinear realization}
Now we are in a position to consider in more detail the condition
(\ref{RTcon}). We have already mentioned that this requirement
can be treated as the condition of transition to the pure
version of the NR. 
What does it mean from the physical point of view? Here we are
going to get the transparent answer to this question using the
standard technic based on the map of the variables in superspace
\cite{ik,ik1} which establishes the connection between linear and
nonlinear realizations.

\subsection{Quasilinear scalar supermultiplet}
Let us, as in the LR method 
(see Eqs.(\ref{RSSt}) and (\ref{shSFtr})) supply the nonlinear
realizations (\ref{usfNR}) with the Goldstone superfield
associated with the central charge
\begin{eqnarray} \label{shQSFtr}
\hat{\phi}_{\xi}^{\prime}
(\hat{x}^{\prime},\tau^{\prime}, \hat{\omega}^{\prime}) &=&
\hat{\phi}_{\xi}(\hat{x},\tau, \hat{\omega}) \nonumber\\
&& -\frac{\xi}{2}(\varepsilon^{\alpha}\psi_{\alpha}(\hat{x},\tau) -
 \xi^{\alpha}\tau_{\alpha}).
\end{eqnarray}
From this equation it is quite easy to see that the superfield
$\hat{\phi}_{\xi}(\hat{x},\tau, \hat{\omega}) -
(\xi/2)\tau^{\alpha}\hat{\omega}_{\alpha}$ has the same
transformation law
as the LR superfield 
$\phi_{\xi}(x,\tau, \omega)$ and
therefore can be identified with the former
\begin{equation} \label{CCSFS}
\phi_{\xi}(x,\tau, \omega) =
\hat{\phi}_{\xi}(\hat{x},\tau, \hat{\omega}) -
\frac{\xi}{2}\tau^{\alpha}\hat{\omega}_{\alpha}.
\end{equation}
Thus, after performing the change of the variables (\ref{ordhatred}) in the
LR superfields 
 (\ref{shSF}) one gets
\begin{eqnarray}\label{LLt}
\phi_L(x, \tau, \omega) &=& \hat{\phi}_L(\hat{x},\tau, \hat{\omega})
 + \frac{i\xi}{2}\hat{\omega}^{\alpha}\psi_{\alpha}(\hat{x},\tau)\nonumber\\
& -& \frac{\xi}{2}\tau^{\alpha}\hat{\omega}_{\alpha} + \frac{i\xi}{4}\hat{\omega}^2,
\end{eqnarray}
here the chiral superfield of the NR 
is
introduced
\begin{equation} \label{chSFNR}
\hat{\phi}_L(\hat{x},\tau, \hat{\omega}) =
\hat{\phi}_{\xi}(\hat{x},\tau, \hat{\omega}) +
\frac{i\xi}{4}(\tau^2 + \psi^2(\hat{x},\tau)).
\end{equation}
Eqs. (\ref{LLt}) and (\ref{chSFNR}) together with
(\ref{ordhatred}) establish the required interrelations between
the superfields of linear and nonlinear realizations describing
the
$4D$ supermembrane. Recall, however, that when deriving these
relations we introduced an additional superfield
$\psi_{\alpha}(\hat{x},\tau)$, which is a new independent superfield
unless it is canonically related to
the analogous Goldstone superfield of the LR method 
$\Lambda_{\alpha}(x,\tau)$. To
avoid the doubling of the Goldstone degrees of freedom in our
approach let us study the transformation laws of the real and
imaginary parts of the superfield
$\hat{\phi}_{\xi}$. As it follows from the eq.(\ref{CCSFS}) its
real part does not transform at all because of the relation
\begin{equation} \label{ReRe}
{\mbox{Re}} \phi_{\xi}(x,\tau, \omega) = {\mbox{Re}} \hat{\phi}_{\xi}(\hat{x},\tau, \hat{\omega})
\end{equation}
and the transformations (\ref{shSFtr}). Hence, all the
$N=1$-components in the $\hat{\omega}$-decomposition
\begin{equation} \label{Redec}
{\mbox{Re}} \hat{\phi}_{\xi} = \hat{A} - \frac{i\xi}{4}(\tau^2 + \psi^2) +
i\hat{\omega}^{\alpha}\hat{\Lambda}_{\alpha} + \frac{i}{4}\hat{\omega}^2\hat{F},
\end{equation}
which follows from the Eq.(\ref{chSFNR}) and the corresponding decompositions of
chiral superfield
\begin{eqnarray} \label{chSFcomp}
{\mbox{Re}} \hat{\phi}_L &=& \hat{A} + i\hat{\omega}^{\alpha}\hat{\Lambda}_{\alpha} -
\frac{i}{4}\hat{\omega}^2\hat{F}, \\
{\mbox{Im}} \hat{\phi}_L &=& \hat{B} + i\hat{\omega}^{\alpha}\hat{\Sigma}_{\alpha} -
\frac{i}{4}\hat{\omega}^2\hat{G}, \nonumber
\end{eqnarray}
are transformed {\it independently~} of each other. This allows
one to put any of them equal to zero without the violation of the
covariant properties of the theory. For instance, one can check
that the constraint
\begin{equation} \label{fst}
\hat{\Lambda}_{\alpha}=0
\end{equation}
makes equal the number of the Goldstone degrees of freedom in the
two sides of the eq. (\ref{CCSFS}). To see this let us substitute
the eq. (\ref{LLt}) into the chirality condition (\ref{shcd}).
Using the transformations of the vector and spinor covariant
derivatives
\begin{eqnarray} \label{LRNRder}
 D^{(\tau)}_{\alpha} &=& {\cal D}_{\alpha}
 + \frac{i}{4}(\hat{\omega}^{\rho}{\cal D}_{\alpha}\psi^{\sigma} +
 \rho \leftrightarrow \sigma)\partial_{\rho\sigma} \nonumber\\
 &-& {\cal D}_{\alpha}\psi^{\rho}\partial_{\rho}^{(\hat{\omega})},\\
 \partial_{\alpha\beta} &=& X^{-1\rho\sigma}_{\alpha\beta}({\cal D}_{\rho\sigma} - {\cal
 D}_{\rho\sigma}\psi^{\gamma}\partial_{\gamma}^{(\hat{\omega})}), \nonumber\\
  X^{~~\rho\sigma}_{\alpha\beta} &=& \delta^{\rho\sigma}_{\alpha\beta} +
 \frac{i}{4}(\hat{\omega}^{\rho}{\cal D}_{\alpha\beta}\psi^{\sigma} +
 \rho \leftrightarrow \sigma), \nonumber\\
 D^{(\omega)}_{\alpha} &=& \partial_{\alpha}^{(\hat{\omega})} +
 \frac{i}{2}\hat{\omega}^{\beta}\partial_{\alpha\beta},~~~
 \partial_{\alpha}^{(\hat{\omega})} =
 \frac{\partial}{\partial{\hat{\omega}^{\alpha}}}
 \nonumber
\end{eqnarray}
where ${\cal D}_{\alpha}$ and ${\cal D}_{\alpha\beta}$ are
defined in (\ref{NRent}), (\ref{NRenti}), and taking into account
eqs. (\ref{chSFcomp}) one can
solve the constraints (\ref{shcd}) in terms of fields of the NR method. 
The final form of the solutions is rather complicated but it is
significantly simplified when the restriction (\ref{fst}) together
with the covariant constraint
\begin{equation} \label{scd}
\hat{A}_{\xi} \equiv \hat{A} - \frac{i\xi}{4}(\tau^2 + \psi^2) = 0,
\end{equation}
are imposed. In this case we arrive at the following form of the quasilinear
chiral superfield of the NR 
\[ {\mbox{Re}} \hat{\phi}_L = \frac{i\xi}{4}(\tau^2 + \psi^2) +
\frac{i\xi}{2}\hat{\omega}^2\frac{1}{1 +
\frac{1}{2} {\cal D}^{\alpha} \psi^{\beta}{\cal D}_{\alpha} \psi_{\beta}}, \]
\begin{eqnarray} \label{QLSSF}
 {\mbox{Im}} \hat{\phi}_L = \hat{B} +
\frac{i\xi}{4}\hat{\omega}^2\frac{{\cal D}^{\gamma} \psi_{\gamma}}{1 +
\frac{1}{2} {\cal D}^{\alpha} \psi^{\beta}{\cal D}_{\alpha}
\psi_{\beta}}.
\end{eqnarray}
Note that in eqs. (\ref{QLSSF}) superfields $\psi_{\alpha}$ and
$\hat{B}$ are not independent but express through the scalar
Goldstone superfield $Q$
\begin{equation} \label{psiQ}
\psi_{\alpha} = i\xi^{-1}{\cal D}_{\alpha}Q,\quad \hat{B} = Q +
\frac{i\xi}{2}\tau^{\beta}\psi_{\beta},
\end{equation}
associated with the central charge in the NR 
\cite{ikr}. This superfield is canonically equivalent to the superfield $B$
of LR 
owing to the Eqs.(\ref{ordhatred}) and (\ref{chSFNR}).
For instance putting the first of them onto the surface
$x^{\alpha\beta}=\hat{x}^{\alpha\beta}$, $\omega^{\alpha}=\psi^{\alpha}$ one obtains
\begin{equation}  \label{AhatA}
\hat{A}=A+i\psi \Lambda_{\alpha}+\frac{1}{4} \psi^2 {D^{\tau\alpha}D^{\tau}_{\alpha}} A,
\end{equation}
where $\hat{A}$ and $A$ are defined in (\ref{scd}) and (\ref{A}). One can checks
that the  equation (\ref{AhatA}) has the following well-known solution \cite{ikr,pst}
\begin{eqnarray} \label{psilambda}
\psi^\alpha&=&\frac{\Lambda^\alpha}{\xi+\frac{1}{2} {D^{\tau\alpha}D^{\tau}_{\alpha}} M},\\
M &\equiv& \frac{\Lambda^2}{{\xi+\sqrt{\xi^2+{D^{\tau\alpha}D^{\tau}_{\alpha}} \Lambda^2}}}.\nonumber
\end{eqnarray}
Substituting this solution into the corresponding equation for the second
scalar superfield
\begin{equation} \label{BhatB}
\hat{B} =  B + \psi^{\alpha} D^{(\tau)}_{\alpha} A +
 \frac{1}{4} \psi^2 D^{(\tau) \alpha}D^{(\tau)}_{\alpha}B
\end{equation}
we get one more relation
\begin{equation} \label{QB}
Q = B - \frac{3}{4}\frac{M~D^{(\tau) \alpha}D^{(\tau)}_{\alpha}B}{{\xi+\sqrt{\xi^2+{D^{\tau\alpha}D^{\tau}_{\alpha}} \Lambda^2}}},
\end{equation}
which proves the equivalence of linear and nonlinear parameterizations
of theory.

In this connection one should note that after imposing the
constraint (\ref{scd}) all the components of the superfield
$\hat{\phi}_L$ are expressed through the scalar Goldstone superfield
of the central charge only. This superfield  (which we call
{\it quasilinear~superfield}) is associated with the limit of the pure nonlinear
realization. From the physical point of view the latter describes
the models in which the mass of the massive mode of supermembrane
tends to infinity. Thus we see that eq.(\ref{scd}) describes the
constraint which removes the contribution of the corresponding
degree of freedom in this limit in its manifestly covariant and
physically transparent form.  The remarkable feature of this
constraint is that it together with (\ref{fst}) automatically
reproduces the nilpotency constraint {\it a l\'{a}} Ro\v{c}ek and
Tseytlin (\ref{RTcon}).

\subsection{Quasilinear vector supermultiplet}

In the case of a $D2$-brane the situation becomes substantially
more complicated because of  another linear supermultiplet which
involves a corresponding worldvolume Goldstone superfield
\cite{ikr,z}.

Recall that this model is described by a real linear superfield
satisfying the {\it deformed~} constraints \cite{z}
\begin{eqnarray} \label{Wxi}
&(D^{(\tau)})^2W_{\xi} - (D^{(\omega)})^2W_{\xi} = 2i\xi,&\nonumber\\
&D^{(\tau)\alpha}D^{(\omega)}_{\alpha}W_{\xi} = 0.&
\end{eqnarray}
The solution of this constraints can be written in the form
\begin{equation} \label{W}
W_{\xi} = W + \frac{i\xi}{2}\omega^2,
\end{equation}
where $W$ is restricted by ordinary constraints with $\xi = 0$.
To be able to derive the corresponding Born-Infeld action we
should impose one more constraint
\begin{equation} \label{RTcons}
W_{\xi}^2 = 0,
\end{equation}
which as before, allows one to avoid the massive degree of
freedom. If we would like to know what does it mean from the
point of view of
the NR method 
we must perform the transformations (\ref{ordhatred})
in the shifted superfield
\begin{equation} \label{hatW}
W_{\xi}(x, \tau, \omega) =
\hat{W}_{\xi}(\hat{x}, \hat{\tau}, \hat{\omega}).
\end{equation}
Now let us consider the following {\it Ansatz~} of quasilinearity
\begin{equation} \label{Anz}
\hat{W}_{\xi}(\hat{x}, \hat{\tau}, \hat{\omega}) =
- \frac{i}{4}\hat{\omega}^2\hat{F},
\end{equation}
which solves exactly the constraint (\ref{RTcons}). Substituting Eq.(\ref{Anz})
into the irreducibility conditions (\ref{Wxi}) and solving it one discovers
\begin{eqnarray} \label{QLLSF}
\hat{W}_{\xi}(\hat{x}, \hat{\tau}, \hat{\omega}) &=&
- \frac{i\xi}{2}\hat{\omega}^2\frac{1}{1 -
\frac{1}{2} {\cal D}^{\alpha} \psi^{\beta}{\cal D}_{\alpha} \psi_{\beta}},\nonumber\\
{\cal D}^{\alpha} \psi_{\alpha} &=& 0.
\end{eqnarray}
Thus we get the $D2$-brane quasilinear superfield (\ref{QLLSF}).
Note that it differs from the analogous superfield of the
supermembrane (\ref{QLSSF}) not only by the sign in the
denominator of the highest  component $\hat{F}$ but also because
it satisfies essentially different constraints. As in the case of
supermembrane this constraint can be rewritten in terms of the
Goldstone
superfield of the underlying LR method 
\begin{equation} \label{DL}
D^{(\tau)\alpha}\Lambda_{\alpha}= 0.
\end{equation}
The latter follows immediately from eqs. (\ref{Wxi}) and
(\ref{W}) applied to the $N=1$ decomposition of the superfield $W$
\begin{equation} \label{lSFcomp}
W = A + i\omega^{\alpha}\Lambda_{\alpha} + \frac{i}{4}\omega^2 F.
\end{equation}
In eq. (\ref{lSFcomp}) the component fields $A$ and $F$ are chosen
to be real. As it was shown in \cite{ikr} the constraint
(\ref{DL}) ensures between the supermembrane and the $D2$-brane on
the level of the corresponding superfield actions.

\section{Actions}
Now we would like to discuss the problem of actions. In the
framework of the linear realizations the actions for the
supermembrane and the
$D=2$-brane were constructed
in \cite{ikr}. However the corresponding action for the
superfields of the nonlinear realization is known only for the
case of the supermembrane
\cite{ikr,pst}.
Having at our disposal the quasilinear superfields (\ref{QLSSF})
and (\ref{QLLSF}) we can get both  forms of the actions starting
straightforwardly from the underlying actions of the LR formulation: 
\begin{eqnarray} \label{actsLR}
S_{SM} &=& -
\frac{i}{4}\int d^3x d^2\tau d^2\omega ~ \omega^2 {\mbox{Re}} \phi_{\xi}, \\
S_{D2} &=& -
\frac{i}{4}\int d^3x d^2\tau d^2\omega ~ \omega^2 W_{\xi}. \nonumber
\end{eqnarray}
Making in (\ref{actsLR}) the change of coordinates $z =
\{x,
\tau,
\omega
\}$
$\Rightarrow$ $\hat{z} = \{\hat{x}, \hat{\tau}, \hat{\omega} \}$ one gets
$$S_{SM} = -
\frac{i}{4}\int d^3\hat{x} d^2\hat{\tau} d^2\hat{\omega}~
\mbox{sdet}(\frac{\partial z}{\partial \hat{z}})(\hat{\omega} + \psi)^2
{\mbox{Re}} \hat{\phi}_{\xi},$$
$$S_{D2} = -
\frac{i}{4}\int d^3\hat{x} d^2\hat{\tau} d^2\hat{\omega}~
\mbox{sdet}(\frac{\partial z}{\partial \hat{z}})(\hat{\omega} + \psi)^2
\hat{W}_{\xi}.$$
Despite of its complexity Though these actions look rather
complicated  they can be represented in a compact explicit form
due to a very special structure of the quasilinear superfield
(\ref{QLSSF}) and (\ref{QLLSF})
\begin{eqnarray} \label{ffacts}
S_{SM} &=& -
\frac{\xi}{2}\int d^3\hat{x} d^2\hat{\tau}\frac{\psi^2}{1 +
\frac{1}{2} {\cal D}^{\alpha} \psi^{\beta}{\cal D}_{\alpha} \psi_{\beta}}, \\
S_{D2} &=& -
\frac{\xi}{2}\int d^3\hat{x} d^2\hat{\tau}\frac{\psi^2}{1 -
\frac{1}{2} {\cal D}^{\alpha} \psi^{\beta}{\cal D}_{\alpha} \psi_{\beta}}.
\nonumber
\end{eqnarray}
Note that in these actions the superfields $\psi_{\alpha}$ are
constrained by the conditions (\ref{psiQ}) and (\ref{QLLSF}),
respectively.

\section{Conclusion}

In the present talk we have described the connection between the
linear and nonlinear realizations of models with partially broken
global $N=2$, $D=3$ supersymmetry. With the example of the
$4D$ supermembrane and the
$D2$-brane we have shown that these realizations are canonically equivalent
to each other off the mass shell due to the existence of the map
transformations relating the corresponding Goldstone superfields.
It is quite evident, however,  that this approach is not
restricted to these canonical models only. It can be successfully
applied to any $PBGS$ models which allow for a superspace
formulation. But the most essential feature of this approach is
that it opens a new possibilities of investigation of models with
partially broken
{\bf local} supersymmetries.

We hope to consider these problems in detail in the framework of
the nonlinear realization of $PBLS$ gauge theory in forthcoming
publications.

We are indebted to E. Ivanov, S. Krivonos and D. Sorokin for asking
the questions which led to this work.

\end{document}